\documentclass[twocolumn]{autart}    % Enable this line and disable the
\usepackage{amsmath,amssymb,amsfonts}
\usepackage{lipsum,cuted}
\usepackage{dsfont}
\usepackage{graphicx}
\usepackage{algorithm,algorithmic}
\usepackage{color}
\usepackage{cite}
\usepackage{subcaption}
\usepackage{epstopdf}
\usepackage{bm}
\usepackage{mleftright}
\usepackage{mathrsfs}
\usepackage{stmaryrd}
\usepackage{apacite}

\newtheorem{theorem}{Theorem}

\newtheorem{lemma}{Lemma}

\newtheorem{definition}{Definition}

\newcommand{\black}[1]{{\color{black}{#1}}}

\def\Re{\mbox{\rm Re}}
\def\Im{\mbox{\rm Im}}

\def\rank{\mbox{\rm rank}}

\def\mr+{\mbox{$\text{mr}_{+}$}}

\graphicspath{{./fig/}}
\begin{document}

\begin{frontmatter}
%\runtitle{Insert a suggested running title}  % Running title for regular
                                              % papers but only if the title
                                              % is over 5 words. Running title
                                              % is not shown in output.

\title{
Converse negative imaginary theorems}

\thanks[footnoteinfo]{This work was supported in part by the Engineering and Physical Sciences Research Council (EPSRC) [grant number EP/R008876/1], and by the National Natural Science Foundation of China [grant number 62103303]. All research data supporting this publication are directly available within this publication. For the purpose of open access, the authors have applied a Creative Commons Attribution (CC BY) licence to any Author Accepted Manuscript version arising.}

\author[SZ]{Sei Zhen Khong}\ead{szkhongwork@gmail.com},   
\author[TJ,ost]{Di Zhao}\ead{dzhao925@tongji.edu.cn},    % Add the   
\author[Man]{Alexander Lanzon}\ead{Alexander.Lanzon@manchester.ac.uk}

\thanks[ost]{Corresponding author.}
\address[SZ]{Independent Researcher}  
\address[TJ]{Department of Control Science and Engineering \& Shanghai Institute of Intelligent Science and Technology, Tongji University, Shanghai, China}
\address[Man]{Control Systems Centre, Department of Electrical and Electronic Engineering, School of Engineering, University of Manchester, M13 9PL Manchester, UK}

\vspace{-5pt}
                             % keyword list or with the
                                          % help of the Automatica
                                          % keyword wizard

\begin{abstract}                          
Converse negative imaginary theorems for linear time-invariant systems are derived. In particular, we provide necessary and sufficient conditions for a feedback system to be robustly stable against various types of negative imaginary (NI) uncertainty. \black{Uncertainty classes of marginally stable NI systems and stable strictly NI systems with restrictions on their static or instantaneous gains are considered.} It is shown that robust stability against the former class entails the strictly NI property, \black{whereas the latter class entails the NI property}. We also establish a non-existence result that no stable system can robustly stabilise all marginally stable NI uncertainty, thereby showing that the uncertainty class of NI systems is too large as far as robust feedback stability is concerned, thus justifying the consideration of subclasses of NI systems with constrained static or instantaneous gains.
\end{abstract}

\end{frontmatter}

%%%%%%%%%%%%%%%%%%%%%%%%%%%%%%%%%%%%%%%%%%%%%%%%%%%%%%%%%%%%%%%%%%%%%%%%%%%%%%%%
\section{Introduction}

In the field of robust control, converse results are intimately tied to the conservativity of robust feedback stability conditions --- they show that such conditions are not conservative if robust stability against particular uncertainty classes is \black{required}. These results were first derived for the small-gain theorem on linear time-invariant (LTI) systems and $\mu$-analysis; see~\cite[Theorem 9.1]{ZDG96}, \cite{doyle1991structured} and the references therein. They were further explored within the setting of passivity, and more generally, integral quadratic constraints in~\cite{khong2021TAC, khong2022TAC}. Interesting robotic applications of converse passivity theorems were described in~\cite{colgate1988robust, stramigioli2015energy, khong2018Auto}. Converse results are also prevalent in the literature on the graph topology~\cite{vidyasagar2011control, georgiou1990optimal, vinnicombe1993frequency,qiu1992feedback,di2020tac,di2021auto}. Recently, attempts to obtain converse results for Lurye systems involving monotone nonlinearity with the so-called Zames-Falb multipliers have been made in~\cite{khong2021necessity, su2023necessity}, with significant breakthrough achieved in~\cite{kharitenko2022exactness}.  The importance of converse results is hence self-explanatory given their ubiquity in the literature.

This paper is concerned with converse results for linear time-invariant negative imaginary (NI) systems. \black{NI} systems theory was first proposed by \cite{Lanzon2008TAC} and was originally motivated in part by robust vibration control of flexible systems~\cite{Petersen2010MCS,Ian2016AnnualRev}. NI theory was subsequently extended to include imaginary-axis poles~\cite{TAC-10_Xiong-Petersen-Lanzon}, free-body dynamics \cite{Ian2014TAC}, \black{irrational and improper systems~\cite{ferrante2013Auto,ferrante2016TAC}}, \black{non-proper systems \cite{LIU2016SCL}, state-space symmetric systems \cite{LIU2019auto}}, discrete-time systems~\cite{AUTOMATICA-17_Ferrante-Lanzon-Ntogramatzidis,liu18,discrete_time_NI2021IJC,LCSS-22_Bhowmick-Lanzon}, nonlinear and time-varying systems~\cite{kurawa2020negative,Petersen2018CDC}, etc. NI theory offers a complementary robust analysis and \black{synthesis framework \cite{BHOWMICK2020auto}} to passivity and small-gain techniques. The NI notion is closely related to counterclockwise I/O dynamics~\cite{angeli2006CCW}. NI theory is an energy-based technique, akin to passivity theory, with connections to Hamiltonian systems~\cite{schaft16}, dissipativity~\black{\cite{bhowmick2019cdc,lanzon2023TAC}} and integral quadratic constraints~\cite{KhongPR17, zhao2022frequency, khong2023feedback}. 

\black{Many practical systems possess NI properties.} Inertial systems, e.g. robotic manipulators, large space structures, unmanned aerial vehicles, mobile robots, etc., driven via force (or torque) actuation and producing a colocated linear (or angular) displacement output possess \black{NI} dynamics arising from physical considerations~\cite{Petersen2010MCS}. A key benefit of \black{NI} theory is that robust stability against a class of physically motivated and physically interpretable uncertainty can be specified in terms of a simple steady-state (i.e. static) gain condition of the open-loop systems~\cite{Lanzon2008TAC}. These developments have enabled NI theory to find interesting applications in many fields including multi-agent systems~\cite{AUTOMATICA-15_Wang-Lanzon-Petersen,AUTOMATICA-20_Skeik-Lanzon} and nanopositioning control~\cite{reza22,TMECH-14_Mabrok-Kallapur-Petersen-Lanzon}, to mention a few.

This paper shows that there exists no stable controller that robustly stabilises all NI uncertain plants with possible poles on the imaginary axis, meaning that the full NI uncertainty class is too large for robust stability with stable controllers. Correspondingly, we derive converse results for numerous different NI uncertainty classes wherein the static and/or instantaneous gains of the systems are restricted. This is possible because some partial knowledge (e.g. a bound) of the static or the instantaneous gain is often available. For this endeavour, new stability results for the feedback interconnection of an NI system and a strictly negative imaginary (SNI) system are established. \black{Two classes of uncertainty are the main focus in this study --- marginally stable NI uncertainty and stable SNI uncertainty.} In particular, we establish that in order to robustly stabilise a class of NI plants with constrained static and/or instantaneous gains, a controller must necessarily satisfy a certain NI property. In other words, the NI property is nonconservative as far as robust feedback stabilisation against NI uncertainty is concerned.

The paper has the following structure. The notation of the paper and preliminaries on systems theory are provided in Section~\ref{sec:notation}. Some new direct NI results are derived in Section~\ref{directNIresults} as they will be needed in the rest of the paper.  In Section~\ref{sec: converse}, various converse results for NI systems are derived. The paper is concluded in Section~\ref{sec: con}.

% \black{An arXiv version of this paper is given in \cite{khong2023converseNI}, where we put several full proofs for theorems in this study due to the space limitation. }
\section{Notation and preliminaries}\label{sec:notation}

\subsection{Basic notation}

Let $\mathbb{F} = \mathbb{R}$ or $\mathbb{C}$ be the real or complex field,  and $\mathbb{F}^n$ be the linear space of $n$-tuples of $\mathbb{F}$ over the field $\mathbb{F}$. The real and imaginary parts of a complex number $s\in\mathbb{C}$ are denoted by $\Re(s)$ and $\Im(s)$, respectively, and its conjugate by $\bar{s}$. The determinant of a matrix $A\in\mathbb{F}^{n\times n}$ is denoted by $\det(A)$, the rank by $\rank(A)$, the transpose $A^T$, the complex conjugate transpose $A^*$, and the singular values
$\bar{\sigma}(A)= \sigma_1(A) \geq\sigma_{2}(A)\geq \cdots \geq \sigma_{n}(A)=\underline{\sigma}(A)$. When eigenvalues of $A$ are real, denote its largest and smallest eigenvalues by $\bar{\lambda}(A)$ and $\underline{\lambda}(A)$, respectively. 
% We say $\lambda$ is a normal eigenvalue of $A\in\mathbb{C}^{n\times n}$ if every eigenvector corresponding to $\lambda$ is orthogonal to any other eigenvectors of $A$. 
For $x,y \in \mathbb{F}^n$, the inner product is denoted as $\langle x,y\rangle=x^*y$, and the Euclidean norm as $|x|=\sqrt{\langle x,x\rangle}$. The identity matrix and zero matrix in $\mathbb{C}^{n\times n}$ are respectively denoted as
$I_n$ and $0_n$. 
% The Moore-Penrose pseudo-inverse of a matrix $A\in\mathbb{C}^{n\times m}$  is denoted by $A^{\dagger}$. 
A matrix is said to be Hurwitz if all of its eigenvalues are located on the open left-half complex plane. \black{Given $A\in\mathbb{F}^{n\times n}$, let $A > 0$ ($A \geq 0$) denote $A$ being positive (semi-)definite. }

% Denote the set of absolutely integrable signals by 
% $\mathcal{L}^n_1
% = \{u : [0, \infty) \to \mathbb{R}^n|~ \int_0^\infty |u(t)| \, dt < \infty\}.$
% Denote the set of energy-bounded signals by
% $\mathcal{L}^n_2
% = \{u : [0, \infty) \to \mathbb{R}^n|~ \|u\|_2^2 = \int_0^\infty |u(t)|^2 \, dt < \infty\}.$
% The Fourier transform of $u\in\mathcal{L}^n_2$ is denoted by $\hat{u}$. Given $T\geq 0$, define the truncation operator $\bm{\Gamma}_{T}$ on all signals $u : [0, \infty) \to \mathbb{R}^n$ via
% $(\bm{\Gamma}_{T} u)(t) = 
% u(t)$ when $0\leq t\leq{T}$ and otherwise
% $(\bm{\Gamma}_{T} u)(t)=0$. 
% Denote the extended $\mathcal{L}_2$ space as
% $\mathcal{L}^n_{2e}=\left\{u :[0, \infty) \to \mathbb{R}^n|~ \bm{\Gamma}_{T} u\in\mathcal{L}^n_2,~\forall~T>0\right\}.$

Denote by $\mathcal{L}_\infty$ the Lebesgue space of functions that are essentially bounded on the imaginary axis $j\mathbb{R}$. Denote by $\mathcal{H}_\infty$
the Hardy space of functions that are holomorphic and uniformly bounded on the open right-half complex plane. The $\mathcal{H}_\infty$ norm of a function $G\in\mathcal{H}_\infty^{n\times n}$ is defined as
$$\|G\|_\infty=\sup_{\text{\rm Re}\,s>0}\bar{\sigma}(G(s))=\sup_{\omega\in\mathbb{R}}\bar{\sigma}(G(j\omega)).$$
Denote by $\mathcal{R}^{n \times n}$ the set of proper real-rational transfer functions and $\mathcal{RH}_\infty^{n\times n}$ the set of all real-rational members in $\mathcal{H}_\infty^{n\times n}$. A $G \in \mathcal{R}^{n \times n}$ is said to be stable if $G\in\mathcal{RH}_\infty^{n\times n}$. In what follows, the superscripts in $\mathcal{H}_\infty^{n\times n}$, $\mathcal{R}^{n\times n}$, ... will be omitted when the context is clear.

\subsection{Feedback stability}

	\begin{figure}[hbt]
		\setlength{\unitlength}{1.2mm}
		\begin{center}
			\begin{picture}(50,25)
				\thicklines \put(0,20){\vector(1,0){8}} \put(10,20){\circle{4}}
				\put(12,20){\vector(1,0){9}} \put(21,16){\framebox(8,8){$P$}}
				\put(29,20){\line(1,0){11}} \put(40,20){\vector(0,-1){13}}
				\put(38,5){\vector(-1,0){9}} \put(40,5){\circle{4}}
				\put(50,5){\vector(-1,0){8}} \put(21,1){\framebox(8,8){$C$}}
				\put(21,5){\line(-1,0){11}} \put(10,5){\vector(0,1){13}}
				\put(0,20){\makebox(5,5){$w_1$}} \put(45,0){\makebox(5,5){$w_2$}}
				\put(13,20){\makebox(5,5){$u_1$}} \put(32,0){\makebox(5,5){$u_2$}}
				\put(33,20){\makebox(5,5){$y_1$}} \put(13,0){\makebox(5,5){$y_2$}}
			\end{picture}
			\caption{A feedback system $[P,C]$.} \label{fig:feedback_sys}
		\end{center}
	\end{figure}
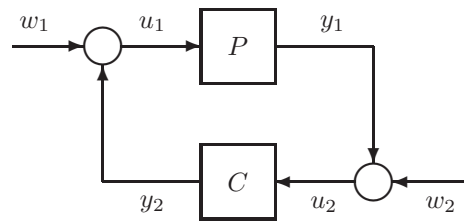
 Denote by $[P,C]$ the positive feedback interconnection of $P\in\mathcal{R}^{n\times n}$ and $C\in\mathcal{R}^{n\times n}$ illustrated in Fig.~\ref{fig:feedback_sys}. The feedback system $[P, C]$ is said to be stable \black{\cite{ZDG96}} if 
 $$\begin{bmatrix}
     I \\ C
 \end{bmatrix}(I-PC)^{-1}\begin{bmatrix}
     I & P
 \end{bmatrix}\in\mathcal{RH}^{2n\times 2n}_\infty.$$

\subsection{Negative imaginary systems}

The following definition of negative imaginary linear time-invariant (LTI) systems is taken from~\cite{Ian2014TAC,Lanzon2017tac}.

\begin{definition}[Negative imaginary systems]\label{def:NI} \cite{Ian2014TAC,Lanzon2017tac}
	A system $G\in\mathcal{R}^{n\times n}$ is said to be negative imaginary (NI) if:
   \begin{enumerate} \renewcommand{\theenumi}{\textup{(\roman{enumi})}}\renewcommand{\labelenumi}{\theenumi}
    \item  $G$ has no poles on the open right half plane;
    \item for all $\omega\in(0,\infty)$ such that $j\omega$ is not a pole of $G$,
      \[
        j\left(G(j\omega)-G(j\omega)^*\right) \geq 0;
      \]
 
    \item for any $\omega_0\in(0,\infty)$, if $j\omega_0$ is a pole of $G$, then it is a simple pole and
      \[
        \lim_{s\to j\omega_0} (s-j\omega_0) j G(s) \geq 0;
      \]
      
    \item if $0$ is a pole of $G$, then $\displaystyle\lim_{s \to 0} s^k G(s) = 0$ for each integer $k \geq 3$ and
      \[
        \lim_{s\to 0} s^2G(s) \geq 0.
      \]
      \end{enumerate}
\end{definition}

A strict subclass of NI systems, termed strictly negative imaginary, was defined in~\cite{Lanzon2008TAC} and is restated here for convenience. Its relation to several other strict subclasses, defined later in the literature, is described in~\cite{lanzon2023TAC}.

\begin{definition}[Strictly NI systems]\label{def:SNI} \cite{Lanzon2008TAC}
	A system $G\in\mathcal{RH}^{n\times n}_\infty$ is said to be strictly negative imaginary (SNI) if 
	\[
    j\left(G(j\omega)-G(j\omega)^*\right) > 0 \quad\forall \omega\in(0,\infty).
    \]
\end{definition}

% Next, we define a new class of output strictly negative imaginary systems, different from the well-known OSNI systems already existing in the literature~\cite{lanzon2023TAC}. The reader will notice that there are two key differences: (i) $\epsilon$ is frequency-dependent, and (ii) the right-hand side of inequality~\eqref{new_OSNI_inequality} contains simply $G(j\omega)^*G(j\omega)$ instead of $\omega\bar{G}(j\omega)^*\bar{G}(j\omega)$ with $\bar{G}(s)=G(s)-G(\infty)$.

% \begin{definition}[Output Strictly NI systems] \label{def:OSNI}
% 	A system $G\in\mathcal{RH}^{n\times n}_\infty$ is said to be output strictly negative imaginary (OSNI) if for all $\omega \in (0, \infty)$, there exists $\epsilon = \epsilon(\omega) > 0$ such that
%  \begin{equation}\label{new_OSNI_inequality}
%  j\left(G(j\omega)-G(j\omega)^*\right) \geq \epsilon G(j\omega)^*G(j\omega).
%  \end{equation}
%  \end{definition}

Next, we provide the standard definition of positive real systems~\cite{brian2006Network} for ease of reference.

\begin{definition}[Positive real systems] \cite{brian2006Network}
A system $G\in\mathcal{RH}^{n\times n}_\infty$ is said to be passive (a.k.a. positive real) if
\[
G(j\omega) + G(j\omega)^* \geq 0 \quad\forall \omega \geq 0,
\]
and output strictly passive if there exists $\epsilon > 0$ such that
\[
G(j\omega) + G(j\omega)^* \geq \epsilon G(j\omega)^*G(j\omega) \quad\forall \omega \geq 0.
\]
\end{definition}

A number of useful results from the literature are collected here because they will be used throughout this paper. Lemma~\ref{lem:propertyNI_zero_infty} gives useful properties on the static (i.e. at zero frequency) and instantaneous (i.e. at infinity frequency) gains of NI systems.

\begin{lemma}\cite[Lemma~2]{Lanzon2008TAC} and~\cite[Lemma~8]{Lanzon2017tac}\label{lem:propertyNI_zero_infty}
Let $G\in\mathcal{R}^{n\times n}$. Then the following statements are true. 
\begin{enumerate}
    \item [{\rm(a)}] If $G$ is NI without poles at origin, then
    \[
    G(0)=G(0)^T, G(\infty)=G(\infty)^T, \text{and } G(0) \geq G(\infty).
    \]
    \item [{\rm(b)}] If $G$ is SNI, then
    $$G(0) > G(\infty).$$
\end{enumerate}
\end{lemma}

Lemma~\ref{lem:stability_WithoutOriginPole} gives necessary and sufficient conditions for the stability of a positive feedback interconnection of an NI system without poles at the origin with an SNI system.

\begin{lemma}\cite[Theorem~9]{Lanzon2017tac}\label{lem:stability_WithoutOriginPole}
Let $P$ be NI without poles at origin and $C$ be SNI. Then $[P,C]$ is stable if and only if
\begin{enumerate}
    \item [{\rm(a)}] $I-P(\infty)C(\infty)$ is nonsingular;
    \item [{\rm(b)}] $\bar{\lambda}\left((I-P(\infty)C(\infty))^{-1}(P(\infty)C(0)-I)\right)<0$;
    \item [{\rm(c)}] $\bar{\lambda}\left((I-C(0)P(\infty))^{-1}(C(0)P(0)-I)\right)<0$.
\end{enumerate}
\end{lemma}

Lemma~\ref{lem:stability_WithoutOriginPole2} gives alternative necessary and sufficient conditions to those provide in Lemma~\ref{lem:stability_WithoutOriginPole}.

\begin{lemma}\cite[Theorem~14]{Lanzon2017tac}\label{lem:stability_WithoutOriginPole2}
Let $P$ be NI without poles at origin and $C$ be SNI. Then $[P,C]$ is stable if and only if
\begin{enumerate}
    \item [{\rm(a)}] $I-P(\infty)C(\infty)$ is nonsingular;
    \item [{\rm(b)}] $\bar{\lambda}\left((P(0)C(\infty)-I)(I-P(\infty)C(\infty))^{-1}\right)<0$;
    \item [{\rm(c)}] $\bar{\lambda}\left((C(0)P(0)-I)(I-C(\infty)P(0))^{-1}\right)<0$.
\end{enumerate}
\end{lemma}

Lemma~\ref{lem:stability_WithOriginPole} gives necessary and sufficient conditions when the NI system is allowed to have poles at the origin. Further discussion on Lemmas~\ref{lem:propertyNI_zero_infty} to~\ref{lem:stability_WithOriginPole} can be found in~\cite{Lanzon2017tac}.

\begin{lemma}\cite[Theorem~24]{Lanzon2017tac}\label{lem:stability_WithOriginPole}
Let $P$ be NI and $C$ be SNI. Let $\Psi<0$ be such that $\bar{\lambda}(P(\infty)\Psi)<1$. Then $[P,C]$ is stable if and only if
\begin{enumerate}
    \item [{\rm(a)}] $I-P(\infty)C(\infty)$ is nonsingular;
    \item [{\rm(b)}] $\bar{\lambda}\left((I-P(\infty)C(\infty))^{-1}(P(\infty)C(0)-I)\right)<0$;
    \item [{\rm(c)}] $\bar{\lambda}\left(\lim_{s\to 0}(I-\Psi P(\infty))(I-C(s)P(\infty))^{-1}\times\right.$\\
    $\left.\hspace{50pt}(C(s)P(s)-I)(I-\Psi P(s))^{-1}\right)<0$.
\end{enumerate}
\end{lemma}

The following result is taken from~\cite[Proposition~1]{khong2022TAC}, which in turn relies on~\cite[Proposition~1]{khong2021TAC}. 

\begin{lemma}\label{lem:converse_PR}
Let $K\in\mathcal{RH}_\infty^{n\times n}$. Then $[N,-K]$ is stable for all positive real $N\in\mathcal{RH}_\infty^{n\times n}$ if and only if $K$ is output strictly passive. In particular, if $K(j\omega_0) + K(j\omega_0) \ngeq \epsilon K(j\omega_0)^*K(j\omega_0)$ for some $\omega_0 \in [0, \infty]$ and all $\epsilon > 0$, then there exists positive real $N\in\mathcal{RH}_\infty^{n\times n}$ such that $\det(I + K(j\omega_0)N(j\omega_0)) = 0$.
\end{lemma}

\section{Direct Negative Imaginary results}\label{directNIresults}

We first derive some direct results on the stability of positive feedback interconnections of NI systems in this section.

The following theorem provides new necessary and sufficient conditions for the feedback stability of an NI system in positive feedback with an SNI system using a continuous deformation.

\begin{theorem}\label{thm:stabilty_determinant_iff_conditions}
    Suppose $P$ is NI without poles at the origin and $C$ is SNI. Then, the following three statements are equivalent:
    \begin{enumerate}
        \item[{\rm (a)}] $[\tau P, C]$ is stable for all $\tau\in[0,1]$;
        \item[{\rm (b)}] for all $\tau\in[0,1]$,
            \begin{align*}
                &\det(\tau P(\infty)C(\infty)-I) \neq 0, \\
                &\det(\tau P(0)C(\infty)-I) \neq 0 \text{ and} \\
                &\det(\tau P(0)C(0)-I) \neq 0;
            \end{align*}
        \item[{\rm (c)}] for all $\tau\in[0,1]$,
            \begin{align*}
                &\det(\tau P(\infty)C(\infty)-I) \neq 0, \\
                &\det(\tau P(\infty)C(0)-I) \neq 0 \text{ and} \\
                &\det(\tau P(0)C(0)-I) \neq 0.
            \end{align*}
    \end{enumerate}
\end{theorem}
\begin{pf}[(a) $\Leftrightarrow$ (b)]
  Note that $\tau P(s)$ is NI without poles at the origin for all $\tau\in[0,1]$. Then,
  \begin{align*}
    \forall\tau&\in [0,1], \\
    &\det[\tau P(\infty)C(\infty)-I] \neq 0, \\
    &\det[\tau P(0)C(\infty)-I] \neq 0, \\
    &\det[\tau P(0)C(0)-I] \neq 0. \\
    \Leftrightarrow\;\forall\tau&\in [0,1], \\
    &\det[I-\tau P(\infty)C(\infty)] \neq 0, \\
    &\det([\tau P(0)C(\infty)-I][I-\tau P(\infty)C(\infty)]^{-1}) \neq 0, \\
    &\det([C(0)\tau P(0)-I][I-C(\infty)\tau P(0)]^{-1}) \neq 0.
    \end{align*}
    \begin{align*}
    \Leftrightarrow\;\forall\tau&\in [0,1], \\
    &[I-\tau P(\infty)C(\infty)] \text{ is nonsingular}, \\
    &\overline{\lambda}([\tau P(0)C(\infty)-I][I-\tau P(\infty)C(\infty)]^{-1}) < 0, \\
    &\overline{\lambda}([C(0)\tau P(0)-I][I-C(\infty)\tau P(0)]^{-1}) < 0. \\[0.25\baselineskip]
    &\begin{minipage}{21em}
        [This equivalence is because the eigenvalues of $X=([\tau P(0)C(\infty)-I][I-\tau P(\infty)C(\infty)]^{-1})$ and $Y=([C(0)\tau P(0)-I][I-C(\infty)\tau P(0)]^{-1})$ are real via~\cite[Lemmas~3 and 4]{Lanzon2017tac}, the conditions are obviously satisfied at $\tau=0$, the eigenvalues of $X$ and $Y$ deform continuously in $\tau$, and $\overline{\lambda}(X)$ and $\overline{\lambda}(Y)$ do not touch zero as $\tau$ increases from $0$.]
    \end{minipage} \\
    \Leftrightarrow\;[\tau &P, C]\text{ is stable for all $\tau\in[0,1]$ via Lemma~\ref{lem:stability_WithoutOriginPole2}}.
\end{align*}
This concludes the proof of (a) $\Leftrightarrow$ (b).

The proof of (a) $\Leftrightarrow$ (c) is similar to the proof of (a) $\Leftrightarrow$ (b) but invokes Lemma~\ref{lem:stability_WithoutOriginPole} instead of Lemma~\ref{lem:stability_WithoutOriginPole2}.
$\hfill\qed$
\end{pf}

The following result is a specialisation of Theorem~\ref{thm:stabilty_determinant_iff_conditions} using conditions that are easier to check.

\begin{theorem}\label{thm:stability_max_eigenvalue_sufficient_conditions}
    Suppose $P$ is NI without poles at the origin, $C$ is SNI, and either $P(\infty)\ge 0$ or $C(\infty)\geq 0$. Then, $[P, C]$ is stable if 
    $$\overline{\lambda}[P(0)C(0)]<1~~\text{and}~~\overline{\lambda}[P(\infty)C(\infty)]<1.$$
\end{theorem}

\begin{pf}
 We first proof the case when $P(\infty)\ge 0$.
 By Lemma~\ref{lem:propertyNI_zero_infty} and the suppositions, we know that $C(0) = C(0)^T > C(\infty) = C(\infty)^T$ and $P(0)=P(0)^T\geq P(\infty)=P(\infty)^T \ge 0$. Therefore, via~\cite[Lemma~11]{Lanzon2017tac}, $\lambda_i[P(0)C(0)]\in\mathbb{R}$, $\lambda_i[P(\infty)C(\infty)]\in\mathbb{R}$, and $\lambda_i[P(0)C(\infty)]\in\mathbb{R}$ for all $i$. Then,
    \begin{align*}
        &\det(\tau P(\infty)C(\infty)-I) \neq 0 \quad\forall\tau\in [0,1] \\
        \Leftrightarrow\quad &{\lambda_i}[P(\infty)C(\infty)] \neq \frac{1}{\tau} \quad\forall\tau\in (0,1], i \\
        \Leftrightarrow\quad &\overline{\lambda}[P(\infty)C(\infty)] < 1.
    \end{align*}
 Similarly,
    \begin{align*}
        &\det(\tau P(0)C(0)-I) \neq 0 \quad\forall\tau\in [0,1] \\
        \Leftrightarrow\quad &\overline{\lambda}[P(0)C(0)] < 1
    \end{align*}
 and
    \begin{align*}
        &\det(\tau P(0)C(\infty)-I) \neq 0 \quad\forall\tau\in [0,1] \\
        \Leftrightarrow\quad &\overline{\lambda}[P(0)C(\infty)] < 1.
    \end{align*}
  Now, $\overline{\lambda}[P(0)C(0)]<1 \Rightarrow P(0)^{\frac{1}{2}}C(0)P(0)^{\frac{1}{2}} < I \Rightarrow P(0)^{\frac{1}{2}}C(\infty)P(0)^{\frac{1}{2}} < I \Rightarrow \overline{\lambda}[P(0)C(\infty)] < 1$.
  
  The conclusion then follows via Theorem~\ref{thm:stabilty_determinant_iff_conditions} using its conditions (a) and (b).

  The proof of the case when $C(\infty)\geq 0$ is similar to the above proof but invokes conditions (a) and (c) of Theorem~\ref{thm:stabilty_determinant_iff_conditions},  and exploits $C(0)=C(0)^T > C(\infty)=C(\infty)^T\geq 0$ and $P(0)=P(0)^T\geq P(\infty)=P(\infty)^T$ instead.
$\hfill\qed$
\end{pf}

\section{Converse NI results} \label{sec: converse}

A collection of converse results for robust feedback stability of NI systems is now derived in this section. 

\subsection{Uncertainty with possible poles on the imaginary axis}

In this subsection, we derive converse results for uncertain NI systems with possible poles on the imaginary axis. Uncertain NI systems with possible poles at the origin are derived first, and those without poles at the origin are derived next.

\subsubsection{Uncertainty with possible poles at the origin}

%For convenience, we define the system class $\mathcal{N}_1$ which is a subset of the class of NI systems.
%
%\begin{definition}
%  A system $P\in\mathcal{R}^{n\times n}$ is said to belong to $\mathcal{N}_1$ if $P$ is NI, $P(\infty)=0$ and $P$ has no double poles at the origin.
%\end{definition}

Theorem~\ref{thm: SNI_nec} states that an SNI controller with negative static gain is necessary and sufficient for robust stability of the full class of strictly proper NI systems.

\begin{theorem} \label{thm: SNI_nec}
Let $C\in\mathcal{RH}_\infty^{n\times n}$. Then the following three statements are equivalent:
\begin{enumerate} \renewcommand{\theenumi}{\textup{(\roman{enumi})}}\renewcommand{\labelenumi}{\theenumi}
    \item $[P,C]$ is stable for all strictly proper NI $P\in\mathcal{R}^{n\times n}$; \label{item: SNI_nec_NI}
    \item $[P, C]$ is stable for all strictly proper NI $P\in\mathcal{R}^{n\times n}$ that have no double poles at the origin; \label{item: SNI_nec_N}   %$P\in\mathcal{N}_1$; 
    \item $C$ is SNI and $C(0)< 0$. \label{item: SNI_nec_C}
\end{enumerate}
\end{theorem}

\begin{pf}
That \ref{item: SNI_nec_NI} implies \ref{item: SNI_nec_N} is trivial. That \ref{item: SNI_nec_C} implies \ref{item: SNI_nec_NI} may be established using Lemma~\ref{lem:stability_WithOriginPole} as follows. In particular, since $P(\infty)=0$, $[P,C]$ is stable if statement~(c) of Lemma~\ref{lem:stability_WithOriginPole} holds, i.e.,
% only the 3rd condition in \cite[Theorem~24]{Lanzon2017tac} needs to be verified, i.e., for some $\Psi<0$ it holds $\bar{\lambda}(P(\infty)\Psi)<1$ and
% \begin{multline*}
%     \bar{\lambda}\big(\lim_{s\to 0}[(I-\Psi P(\infty))(I-C(s)P(\infty))^{-1}\\
%     (C(s)P(s)-I)(I-\Psi P(s))^{-1}]\big)<0.
% \end{multline*}
% Note that $P(\infty)=0$, and it then suffices to show 
\begin{align*}
\bar{\lambda}\big(\lim_{s\to 0}[(C(s)P(s)-I)(I-\Psi P(s))^{-1}]\big)<0
\end{align*}
for some $\Psi \in \mathbb{R}^{n \times n}$ such that $\Psi < 0$. 
Note that by \cite[Lemma~20]{Lanzon2017tac}, $P(s)(I-\Psi P(s))^{-1}$ has no pole at origin. By choosing $\Psi=C(0)<0$, we then have 
\begin{align*}
    & \lim_{s\to 0}[(C(s)P(s)-I)(I-\Psi P(s))^{-1}] \\
    = \, & \lim_{s\to 0}[(C(s)P(s) - \Psi P(s) + \Psi P(s) -I)(I-\Psi P(s))^{-1}] \\
    = \, & \lim_{s\to 0}[(C(s) - \Psi)P(s)(I-\Psi P(s))^{-1} - I] \\
    = \, & -I.
\end{align*}
Therefore, we obtain
\begin{align*}
\bar{\lambda}\left(\lim_{s\to 0}[(C(s)P(s)-I)(I-\Psi P(s))^{-1}]\right) 
=-1<0.
\end{align*}
This shows the stability of $[P,C]$, whereby  \ref{item: SNI_nec_NI} holds. 

% Let $P=NM^{-1}$ be a right coprime factorization, where $N,M\in\mathcal{RH}_\infty$. Then
% \begin{multline*}
%     (C(s)P(s)-I)(I-\Psi P(s))^{-1}  \\
%     = (C(s)N(s)-M(s))(M(s)-\Psi N(s))^{-1}. 
% \end{multline*}
% Since $\{N, M\}$ is right coprime and $\Psi \in \mathbb{R}^{n \times n}$ is nonsingular, it follows that
% \[
% \det(M(0) - \Psi N(0)) \neq 0.
% \]
% Therefore, by choosing $\Psi = C(0) < 0$,
% \begin{align*}
% & \lim_{s \to 0} (C(s)N(s)-M(s))(M(s)-\Psi N(s))^{-1} \\
% = \, & \lim_{s \to 0}(C(s)N(s)-M(s))\lim_{s \to 0}(M(s)-\Psi N(s))^{-1} \\
% = \, & (\Psi N(0)-M(0))(M(0) - \Psi N(0))^{-1} \\
% = \, & -I.
% \end{align*}
% By choosing $\Psi=C(0)<0$, we then have 
% \begin{multline*}%\label{eq:pf_thm1_suf}
% \lim_{s\to 0}C(s)N(s)-M(s) = \Psi N(0)-M(0)\\
% = -\left(\lim_{s\to 0} M(s)-\Psi N(s)\right).
% \end{multline*}
% Moreover, whenever $s=0$ is a transmission zero of $C(s)N(s)-M(s)$, it is the transmission zero of  $\Psi N(s)-M(s)$ in the same direction. Therefore we obtain
% \begin{align*}
% \bar{\lambda}\big(\lim_{s\to 0}[(C(s)P(s)-I)(I-\Psi P(s))^{-1}]\big) 
% =-1<0,
% \end{align*}
% which shows sufficiency. 

We establish that \ref{item: SNI_nec_N} implies \ref{item: SNI_nec_C} via the following three steps. 

\textbf{Step~1:} We show in what follows that for each $\omega\in(0,\infty)$, there exists $\epsilon>0$ (that depends on $\omega$) such that
\begin{equation}\label{eq:OSNI}
 j\left(C(j\omega)-C(j\omega)^*\right) \geq \epsilon C(j\omega)^*C(j\omega).
\end{equation} 
Suppose to the contrapositive that for some $\omega_0\in(0,\infty)$ there exists no $\epsilon>0$ for which \eqref{eq:OSNI} holds. Denote by $K=jC(j\omega_0)$, which satisfies that $K+K^* \not\geq \epsilon K^*K$ $\,\forall\epsilon>0$. By the necessity of Lemma~\ref{lem:converse_PR}, there exists a positive real $N\in\mathcal{RH}^{n\times n}_{\infty}$ such that $\det(I+KN(j\omega_0))=0$. Let
\[
P(s) = \frac{\omega_0}{s}N(s).
\]
It is straightforward to verify that $P$ is NI, $P(\infty)=0$ and $\det(I-P(j\omega_0)C(j\omega_0))=\det(I+KN(j\omega_0))=0$, whereby $[P,C]$ is unstable. Therefore, by contraposition, \eqref{eq:OSNI} is true, which implies that $C\in\mathcal{RH}_\infty^{n\times n}$ is NI as well.  

% \textbf{Step~1:} We show $C$ is NI in what follows. Suppose to the contrapositive that $C$ is not NI. Since $C$ is stable, there exists $\omega_0\in(0,\infty)$ such that $j\left({C}(j\omega_0)-{C}(j\omega_0)^*\right)\not\geq 0$. Denote by $K=jC(j\omega_0)$, which satisfies that $K+K^* \not\geq 0$. By the necessity of the positive-real theorem (see, e.g., \cite[Theorem~1.7.9]{horn1994topics}, \cite{khong2021TAC}), there exists a positive-real $N\in\mathbb{C}^{n\times n}$ such that $\det(I+KN)=0$. By the positive-real interpolation \cite{youla1967interpolation}, there exists a positive real $\tilde{N}\in\mathcal{RH}^{n\times n}$ such that $\tilde{N}(j\omega_0) = N$. Then we construct that
% $$P(s) = \frac{\omega_0}{s}\tilde{N}(s),$$
% which is NI with $P(\infty)=0$ and $\det(I-P(j\omega_0)C(j\omega_0))=\det(I+KN)=0$. Clearly, $[P,C]$ is unstable, and $C$ must be NI by the contraposition. 

\textbf{Step~2:} We show $C(0)< 0$ next. Suppose to the contrapositive that $C(0) \not< 0$. Since $C$ is NI by Step~1, we know that $C(0) = C(0)^T \in \mathbb{R}^{n \times n}$ by Lemma~\ref{lem:propertyNI_zero_infty}(a) and at least one of its eigenvalues is greater or equal to zero, whereby it admits a real Schur decomposition 
$C(0)=UDU^T$ with $U^T U = I$ and $D$ being real diagonal with $[D]_{11}\geq 0$. 

In the case where $[D]_{11} = 0$, let 
\[
M = U\begin{bmatrix}
\alpha & \\
 & 0_{n-1}
\end{bmatrix}U^T \geq 0
\]
for some $\alpha > 0$ to be chosen later and $P=M/s$. Note that $P$ is NI and $P(\infty)=0$. Furthermore, let $C_1(s)= C(s)-C(0) = sE(s)$, where $E \in \mathcal{RH}_\infty$, and $P_1(s) = (I-P(s)C(0))^{-1}P(s)$. It may be observed that
\begin{align}\label{eq:pf_conj1_step2}
    (I-PC)^{-1}P = (I-P_1 C_1)^{-1}P_1;
\end{align}
see the proof of \cite[Lemma~18]{Lanzon2017tac}.
In addition, by choosing $\alpha > 0$ so that $M$ satisfies $\|ME(0)\|<1$ ensures that $I-ME(0)$ is nonsingular and noting that $\lim_{s\to 0} (I-P(s)C(0))^{-1} = I$, we have
\begin{align*}
   \lim_{s\to 0} s(I-P_1 C_1)^{-1}P_1&= (I - ME(0))^{-1}\lim_{s\to 0}sP \\
   & = (I - ME(0))^{-1} M \neq 0.
\end{align*}
Combining this with \eqref{eq:pf_conj1_step2}, we obtain that the origin is a pole of $(I-PC)^{-1}P$, whereby $[P,C]$ is unstable. 

% we have that for $|s|<\epsilon$ being sufficiently small, $C(s)$ can be represented by
% $$C(s) \approx U\begin{bmatrix}
% \alpha s^k & \\
%  & \tilde{D}
% \end{bmatrix}U^T,$$
% for some $\alpha\in\mathbb{R}$ and $k \geq 1$. 
% Construct that $P=M/s$ with
% $$M = U\begin{bmatrix}
% \beta & \\
%  & 0_{n-1}
% \end{bmatrix}U^T \geq 0$$
% such that $\beta> 0$ and $|\alpha\beta|<1$, whereby $\lim_{s\to 0} (1-\alpha\beta s^{k-1}) \neq 0$. 
% One can verify that $P$ is NI with $P(\infty)=0$, and 
% \begin{multline*}
%     \lim_{s\to 0}((I-P(s)C(s))^{-1}P(s)) \\
% = \lim_{s\to 0}\left(U\begin{bmatrix}
% (1-\alpha\beta s^{k-1})^{-1}\beta/s & \\
%  & I_{n-1}
% \end{bmatrix}U^T\right)\\
% =U\begin{bmatrix}
% \infty & \\
%  & I_{n-1}
% \end{bmatrix}U^T,
% \end{multline*}
% whereby $[P,C]$ is unstable. As a summary, by contraposition, we must have that $C(0)< 0$. 

In the case where $[D]_{11} > 0$, let $P=M/(s+1)$ with
$$M = U\begin{bmatrix}
[D]_{11}^{-1} & \\
 & 0_{n-1}
\end{bmatrix}U^T \geq 0.$$ One can verify that $P$ is NI with $P(\infty)=0$, and 
$$\det(I-P(0)C(0)) = 0,$$
whereby $[P,C]$ is unstable. 

Summarising the above cases, we obtain by contraposition that $C(0)< 0$.  

\textbf{Step~3:} Finally, we show that $C$ is SNI. Since \eqref{eq:OSNI} holds, it suffices to show that $C(j\omega)^*C(j\omega)>0$, or $C(j\omega)$ is nonsingular, for any $\omega \in (0,\infty)$. Suppose to the contrapositive that for $\omega_0\in(0,\infty)$, $C(j\omega_0)$ is singular. It then follows from \cite[Theorem~1.6.6]{horn1994topics} and \cite[Property~1.2.5b]{horn1994topics}  that
$$C(j\omega_0) = T\begin{bmatrix} 0 & \\ & D\end{bmatrix}T^*,$$
for some nonsingular $T\in\mathbb{C}^{n\times n}$ and diagonal $D\in\mathbb{C}^{(n-1)\times (n-1)}$. 
Denote by $T^{-*} = \begin{bmatrix}t_1 & t_2 & \cdots & t_n\end{bmatrix}$, and let
$$M = T^{-*}\begin{bmatrix} 1 & \\ & 0_{n-1}\end{bmatrix}T^{-1} = t_1t_1^* \geq 0.$$
Choose $\alpha,\beta\in\mathbb{R}^n$ so that
$
p(s) = \alpha s+ \beta
$
satisfies $p(j\omega_0) = t_1$. Clearly $p(j\omega)p(j\omega)^* \geq 0$ for any $\omega\in (0,\infty)\setminus\{\omega_0\}$ and $p(j\omega_0)p(j\omega_0)^* = M$. 
Let
\[
G(s) = \frac{1}{s^2+\omega_0^2}p(s)p^T(-s)
\]
and note that 
\begin{align*}
j(G(j\omega)-G(j\omega)^*) & = 0~~\text{for all $\omega \in (0, \infty) \setminus \{\omega_0\}$ and}~~\\
\lim_{s\to j\omega_0} (s-j\omega_0) jG(s) & = \frac{j}{2j\omega_0}p(j\omega_0)p(j\omega_0)^* \\
& = \frac{1}{2\omega_0} M \geq 0.
\end{align*}
Hence $G$ is NI. Recall from~\cite[Lemma 7]{Lanzon2017tac} that $G$ is NI if and only if $G - G(\infty)$ is NI and $G(\infty)=G(\infty)^T$. The latter implies that 
\[
jG(\infty) + (jG(\infty))^* = 0, 
\]
and hence by following the arguments in the last part of the proof of~\cite[Proposition 1]{khong2022TAC}, one may show that there exists positive real $N \in \mathcal{RH}_\infty$ such that $N(j\omega_0) = jG(\infty)$. Now define
\[
P(s) = G(s) - G(\infty) + \frac{\omega_0}{s} N.
\]
Evidently, $P$ is NI and $P(\infty) = 0$. Moreover, since $MC(j\omega_0) = 0$, one can verify that
\begin{multline*}\lim_{s\to j\omega_0} (s-j\omega_0)(I-PC)^{-1}P \\
= \lim_{s\to j\omega_0}(s-j\omega_0)P 
= \frac{1}{2j\omega_0} M\neq 0,\end{multline*}
whereby $s=j\omega_0$ is a pole of $(I-PC)^{-1}P$ and  $[P,C]$ is unstable. By contraposition, we must have 
$C(j\omega)$ is nonsingular for any $\omega \in (0,\infty)$. This completes the proof. 
$\hfill \qed$
\end{pf}

The following non-existence result is of interest because it shows that one cannot find a stable controller that achieves robust stability against the full class of NI systems without double poles at the origin. This is because the full NI class without double poles at the origin is too large, not to say the full NI class itself. 

\begin{theorem}
There exists no $C\in\mathcal{RH}_\infty^{n\times n}$ such that $[P,C]$ is stable for all NI $P \in \mathcal{R}^{n \times n}$ without double poles at the origin.
\end{theorem}

%\begin{pf}
%    By Theorem~\ref{thm: SNI_nec}, if such a $C \in \mathcal{RH}_\infty^{n\times n}$ existed, it must be SNI. We show below that $C$ must also satisfy $C(0) = C(\infty) = 0$, which implies that $C$ does not exist because an SNI $C \in \mathcal{RH}_\infty^{n\times n}$ necessarily satisfies $C(0) = C(0)^T > C(\infty) = C(\infty)^T$ by~\cite[Lemma 2]{Lanzon2008TAC}. 
    
%    Suppose to the contrapositive that $C(0) = C(0)^T \neq 0$. Applying Schur decomposition to $C(0)$ yields $C(0) = TDT^T$, where $T \in \mathbb{R}^{n \times n}$ is nonsingular and $D$ is diagonal with $[D]_{11} \neq 0$. Let $P = T^{-T} E T^{-1}$, where $E$ is diagonal and $[E]_{11} = [D]_{11}^{-1}$. Then $P$ is NI and $\det(I - PC(0)) = 0$, whereby $[P, C]$ is unstable. Therefore, $C(0)$ must be $0$. 
    
%    That $C(\infty) = 0$ may be established likewise.
%    $\hfill \qed$
%\end{pf}
%
\begin{pf}
  We prove this via contradiction. Suppose there exists $C\in\mathcal{RH}_\infty^{n\times n}$ such that $[P,C]$ is stable for all NI $P \in \mathcal{R}^{n \times n}$ with at most simple poles at the origin. Then, by Theorem~\ref{thm: SNI_nec}, $C$ is SNI with $C(0)<0$. Choose $P(s)=C(0)^{-1}$. This choice of $P(s)$, which is NI and has no poles at the origin, results in $I-P(s)C(s)$ having a blocking zero at $s=0$. Therefore, for this choice of $P(s)$, $[P,C]$ is not stable, which yields the required contradiction.
$\hfill\qed$
\end{pf}

\subsubsection{Uncertainty without poles at the origin}

When the uncertain NI systems do not have poles at the origin, converse results may be obtained by constraining their static gains. These results are of practical value in applications where some prior knowledge such as a bound on the static gains of the uncertain NI systems is available.

For convenience, we first define the system class $\mathcal{N}_0$.

\begin{definition}
	A system $P\in\mathcal{R}^{n\times n}$ is said to belong to $\mathcal{N}_0$ if $P$ is NI and has no poles at the origin.
\end{definition}

\begin{theorem} \label{thm: SNI_nec_DC_inst}
Let $C\in\mathcal{RH}_\infty^{n\times n}$ and $\gamma > 0$. 
%Then $[P, C]$ is stable for all $P \in \mathcal{N}_0$ satisfying $P(0) \leq \gamma I$ if and only if $C$ is SNI, $C(\infty) \geq 0$ and $C(0) < \frac{1}{\gamma} I$. 
\black{Then the following three statements are equivalent:
\begin{enumerate} \renewcommand{\theenumi}{\textup{(\roman{enumi})}}\renewcommand{\labelenumi}{\theenumi}
    \item $[P, C]$ is stable for all $P \in \mathcal{N}_0$ satisfying $P(0) \leq \gamma I$; \label{item: SNI_nec_DC_1}
    \item $[P, C]$ is stable for all $P \in \mathcal{N}_0$ satisfying $0\leq P(0) \leq \gamma I$; \label{item: SNI_nec_DC_2}   %$P\in\mathcal{N}_1$; 
    \item $C$ is SNI, $C(\infty) \geq 0$ and $C(0) < \frac{1}{\gamma} I$. \label{item: SNI_nec_DC_C}
\end{enumerate}
}
\end{theorem}

\begin{pf}
\black{That \ref{item: SNI_nec_DC_1} implies \ref{item: SNI_nec_DC_2} is trivial. We then show in the following that \ref{item: SNI_nec_DC_C} implies  \ref{item: SNI_nec_DC_1}.} Let $C$ be SNI with $C(\infty)\geq 0$ and $C(0) < \frac{1}{\gamma} I$, and $P\in\mathcal{N}_0$ with $P(0) \leq \gamma I$. By Lemma~\ref{lem:propertyNI_zero_infty}, $C(0)> C(\infty)\geq 0$ and $P(0)\geq P(\infty)$. Hence, $C(0)^{\frac{1}{2}}P(0)C(0)^{\frac{1}{2}} \leq \gamma C(0) < I$, which implies that $\overline{\lambda}[P(0)C(0)]<1$. Similarly, $C(\infty)^{\frac{1}{2}}P(\infty)C(\infty)^{\frac{1}{2}}\leq C(\infty)^{\frac{1}{2}}P(0)C(\infty)^{\frac{1}{2}}\leq \gamma C(\infty) < \gamma C(0) < I$, which implies that $\overline{\lambda}[P(\infty)C(\infty)]<1$. We then conclude that $[P,C]$ is stable via Theorem~\ref{thm:stability_max_eigenvalue_sufficient_conditions}, whereby \ref{item: SNI_nec_DC_1} holds.

\black{Finally, we show that \ref{item: SNI_nec_DC_2} implies \ref{item: SNI_nec_DC_C} via the following three steps. }

  \textbf{Step~1:} We show that $C$ is SNI via a contrapositive argument. Suppose $C$ is not SNI. Then, $\exists\,\omega_0\in(0,\infty), x\in\mathbb{C}^{n}\backslash\{0\}$ such that $x^* j[C(j\omega_0)-C(j\omega_0)^*]x\leq 0$. The inequality is equivalent to $\Im(x^* C(j\omega_0) x) \geq 0$. Choose $\alpha,\beta\in\mathbb{R}^n$ such that $\alpha j\omega_0 +\beta = x$, and define $f(s)=(\alpha s + \beta) \in\mathbb{C}^n$. 
  
  For the case where $x^* C(j\omega_0) x=0$, choose $P(s)=f(s)\frac{\epsilon}{(s^2+\omega_0^2)}f(-s)^T$ with $\epsilon > 0$. 
  % $\epsilon\in \left(0,\frac{\gamma \omega_0^2}{|\beta|^2} \right]$ when $\beta\neq 0$ and $\epsilon\in(0,\infty)$ when $\beta=0$. 
  Note that $P \in \mathcal{N}_0$ and \black{$0 \leq P(0) \leq \gamma I$} for sufficiently small $\epsilon > 0$. Now, 
  \begin{align*}
      &[I-P(s)C(s)]^{-1}P(s) \\
      &\quad= \left[I - \frac{f(s)\epsilon f(-s)^T}{(s^2+\omega_0^2)}  C(s) \right]^{-1} \frac{f(s)\epsilon f(-s)^T}{(s^2+\omega_0^2)}  \\
      &\quad= f(s)[(s^2+\omega_0^2)I - \epsilon f(-s)^T C(s) f(s)]^{-1}\epsilon  f(-s)^T
  \end{align*}
  which clearly shows that $(I - PC)^{-1}P$ has a pole at $j\omega_0$ as $f(-j\omega_0)^TC(j\omega_0)f(j\omega_0) = x^*C(j\omega_0)x = 0$. Thus, $[P,C]$ is not stable for all $P \in \mathcal{N}_0$ satisfying \black{$0 \leq P(0) \leq \gamma I$}.
  %and
  %\begin{align*}
  % &\lim_{s \to j\omega_0} \det[I - P(s)C(s)] \\
  %&\qquad = 1 - \lim_{s \to j\omega_0} \frac{\epsilon}{(s^2+w_0^2)}f(-s)^TC(s)f(s) \neq 0
  %\end{align*}
  %for sufficiently small $\epsilon > 0$ as $f(-j\omega_0)^TC(j\omega_0)f(j\omega_0) = x^*C(j\omega_0)x = 0$. Consequently, $\lim_{s \to j\omega_0} [I - P(s)C(s)]^{-1} %\neq 0$ and $(I - PC)^{-1}P$ has a pole at $j\omega_0$, whereby $[P,C]$ is not stable.

  Next, consider the case where $x^* C(j\omega_0) x\neq 0$. Choose $r>0$ and $\theta\in[0,\pi]$ such that $r e^{j\theta}=x^* C(j\omega_0) x$. Let 
  \begin{align*}
  &a \in 
  \begin{cases}
    \left(\omega_0, \frac{\omega_0 |\beta|}{\sqrt{|\beta|^2-\gamma r}}\right) & \text{when } |\beta|^2>\gamma r, \\
    (\omega_0, \infty) & \text{otherwise},
  \end{cases}\allowdisplaybreaks\\ 
  &b\in\Bigl(\frac{\omega_0 |\beta|}{\sqrt{|\beta|^2+\gamma r}},\omega_0\Bigr),\allowdisplaybreaks\\
  &c \in \left(0, \frac{\gamma r \omega_0}{|\beta|^2}\right),\allowdisplaybreaks\\
  &d \in 
    \begin{cases}
        \Bigl(\omega_0,\frac{\omega_0 |\beta|}{\sqrt{|\beta|^2-\gamma r\operatorname{cos}\theta}}\Bigr) & \text{when } |\beta|^2>\gamma r\operatorname{cos}\theta, \\
        (\omega_0,\infty) & \text{otherwise},
    \end{cases}\allowdisplaybreaks
  \end{align*}
  and $e\in\Bigl(\frac{\omega_0 |\beta|}{\sqrt{|\beta|^2-\gamma r\operatorname{cos}\theta}},\omega_0\Bigr)$ for $\theta \in (\frac{\pi}{2}, \pi)$. Define
  \[ p(s) = 
     \begin{cases}
        \frac{(a^2-\omega^2_0)}{r(s^2+a^2)} & \text{when } \theta = 0, \\  
        \frac{(\omega^2_0-b^2)}{r(s^2+b^2)} & \text{when } \theta = \pi, \\  
        \frac{c\omega_0}{r(s^2+cs+\omega_0^2)} & \text{when }  \theta=\frac{\pi}{2}, \\
        \frac{(d^2-\omega_0^2)\operatorname{sec}\theta}{r\left(s^2+(d^2-\omega_0^2)\frac{\operatorname{tan}\theta}{\omega_0}s+d^2\right)} & \text{when }  \theta\in(0,\frac{\pi}{2}), \\
        \frac{(e^2-\omega_0^2)\operatorname{sec}\theta}{r\left(s^2+(e^2-\omega_0^2)\frac{\operatorname{tan}\theta}{\omega_0}s+e^2\right)} & \text{when }  \theta\in(\frac{\pi}{2},\pi).
     \end{cases} 
   \]
   Evidently, $p(j\omega_0) x^* C(j\omega_0)x = 1$. Observe that $p(s)$ is NI without poles at the origin and has relative degree $2$. Let $P(s)=f(s) p(s) f(-s)^T$ and note that $P\in\mathcal{N}_0$ and satisfies \black{$0 \leq P(0) \leq \gamma I$}. Since, by construction, $[I-P(j\omega_0)C(j\omega_0)]x=0$, whereby $\det[I-P(j\omega_0)C(j\omega_0)] = 0$, it follows that $[P,C]$ is not stable for all $P \in \mathcal{N}_0$ satisfying \black{$0 \leq P(0) \leq \gamma I$}. This completes the proof that $C$ must be SNI.

  Since $C$ is SNI, it follows that $C(0)=C(0)^T\in\mathbb{R}^{n\times n}$ and $C(\infty)=C(\infty)^T\in\mathbb{R}^{n\times n}$ via Lemma~\ref{lem:propertyNI_zero_infty}. 
  
  \textbf{Step~2:} We show that $C(0) < \frac{1}{\gamma} I$ via a contrapositive argument. Suppose that  $C(0) \not < \frac{1}{\gamma} I$. Then, $\overline{\lambda}[C(0)]\geq \frac{1}{\gamma}$. Choose $P(s)=\frac{1}{\overline{\lambda}[C(0)]} I$ and note that $P \in \mathcal{N}_0$ and satisfies \black{$0 \leq P(0) \leq \gamma I$}. Since, by construction, $\det[I-P(0)C(0)] = 0$, it follows that $[P,C]$ is not stable for all $P \in \mathcal{N}_0$ satisfying \black{$0 \leq P(0) \leq \gamma I$}. This completes the proof that $C(0) < \frac{1}{\gamma} I$.

  \textbf{Step~3:} We now show that $C(\infty)\geq 0$ via a contrapositive argument. Suppose $C(\infty)\not\geq 0$. Then, $\underline{\lambda}[C(\infty)]<0$. Choose
  $P(s) = \frac{s}{\underline{\lambda}[C(\infty)](s+1)} I_n$. Note that $P\in\mathcal{N}_0$ and satisfies \black{$0 \leq P(0) \leq \gamma I$}. Since, by construction, $\det[I-P(\infty)C(\infty)] = 0$, it follows that $[P,C]$ is not stable for all $P \in \mathcal{N}_0$ satisfying \black{$0 \leq P(0) \leq \gamma I$}. This completes the proof that $C(\infty)\geq 0$.
  
  \black{The above three steps together show \ref{item: SNI_nec_DC_C}, which completes the proof.} 
  $\hfill \qed$
\end{pf}

The next result restricts the instantaneous gain of the uncertain plants but imposes no limitation on the instantaneous gain of the controller.

\begin{theorem} \label{thm: SNI_nec_DC}
Let $C\in\mathcal{RH}_\infty^{n\times n}$ and $\gamma > 0$. Then $[P, C]$ is stable for all $P \in \mathcal{N}_0$ satisfying $P(\infty) \ge 0$ and $P(0) < \gamma I$ if and only if $C$ is SNI and $C(0) \leq \frac{1}{\gamma} I$. 
\end{theorem}

\begin{pf}
  (Sufficiency) Let $C$ be SNI with $C(0) \leq \frac{1}{\gamma} I$ and $P\in\mathcal{N}_0$ with $P(\infty) \ge 0$ and $P(0) < \gamma I$. By Lemma~\ref{lem:propertyNI_zero_infty}, $P(0)\geq P(\infty) \geq 0$ and $C(0) > C(\infty)$. Thus,
  $P(\infty)^{\frac{1}{2}}C(\infty)P(\infty)^{\frac{1}{2}} \leq P(\infty)^{\frac{1}{2}}C(0)P(\infty)^{\frac{1}{2}} \leq \frac{1}{\gamma} P(\infty) \leq \frac{1}{\gamma} P(0) < I$, which implies that $\bar{\lambda}(P(\infty)C(\infty)) < 1$. Likewise, $P(0)^{\frac{1}{2}}C(0)P(0)^{\frac{1}{2}} \leq \frac{1}{\gamma} P(0) < I$, whereby $\bar{\lambda}(P(0)C(0))<1$. Stability of $[P, C]$ thus follows from Theorem~\ref{thm:stability_max_eigenvalue_sufficient_conditions}.

  (Necessity) The same arguments from the proof for that \ref{item: SNI_nec_DC_2} implies \ref{item: SNI_nec_DC_C} in Theorem~\ref{thm: SNI_nec_DC_inst} are applicable here, since the $P \in \mathcal{N}_0$ constructed therein additionally satisfies $P(0) < \gamma I$ and $P(\infty) \ge 0$.
  $\hfill \qed$
\end{pf}

\subsection{\black{Stable SNI uncertainty}}

In this subsection, converse results involving \black{stable uncertain SNI} systems are obtained.
\begin{theorem} \label{thm: NI_nec}
Let $C\in\mathcal{RH}^{n\times n}_{\infty}$. Then $[P,C]$ is stable for all SNI $P \in \mathcal{RH}_\infty^{n \times n}$ satisfying $P(\infty) \ge 0$ if and only if $C$ is NI and $C(0) \le 0$.  
\end{theorem}

\begin{pf}
(Sufficiency) It follows from Lemma~\ref{lem:propertyNI_zero_infty} that $P(0)>P(\infty) \geq 0$
     and $0 \geq C(0) \geq C(\infty)$. Consequently, one can verify for all $\tau\in[0,1]$, $i\in\{1,2,\dots,n\}$ that
     \begin{align*}
                &\lambda_i(\tau P(\infty)C(\infty)-I) \leq -1, \\
                &\lambda_i(\tau P(\infty)C(0)-I) \leq -1 \text{ and} \\
                &\lambda_i(\tau P(0)C(0)-I) \leq -1,
            \end{align*}
     whereby the stability of $[\tau P,C]$ follows by Theorem~\ref{thm:stabilty_determinant_iff_conditions}. 

 (Necessity)  First, we show that $C$ is NI via a contrapositive argument. Suppose $C$ is not NI. Then, $\exists\,\omega_0\in(0,\infty), x\in\mathbb{C}^{n}\backslash\{0\}$ such that $x^* j[C(j\omega_0)-C(j\omega_0)^*]x< 0$. This equivalent to $\Im(x^* C(j\omega_0) x) > 0$. Choose $\alpha,\beta\in\mathbb{R}^n$ such that $\alpha j\omega_0 +\beta = x$, and define $f(s)=(\alpha s + \beta) \in\mathbb{C}^n$. 
Let $r>0$ and $\theta\in(0,\pi)$ be such that $r e^{j\theta}=x^* C(j\omega_0) x$. Fix $\gamma>0$, and let 
\begin{equation}\label{eq:pf_nec_stableSNI1}
 \begin{aligned}
 &c \in \left(0, \frac{ \gamma r \omega_0}{|\beta|^2}\right),\\
    &d \in 
    \begin{cases}
        \Bigl(\omega_0,\frac{\omega_0 |\beta|}{\sqrt{|\beta|^2-  \gamma r\operatorname{cos}\theta}}\Bigr) & \text{when } |\beta|^2> \gamma r\operatorname{cos}\theta, \\
        (\omega_0,\infty) & \text{otherwise},
    \end{cases}\\
  &\text{and}~e\in\Bigl(\frac{\omega_0 |\beta|}{\sqrt{|\beta|^2-\gamma  r\operatorname{cos}\theta}},\omega_0\Bigr) ~\text{when}~\theta \in (\frac{\pi}{2}, \pi).
   \end{aligned}  
\end{equation}
  Define
  \begin{equation}\label{eq:pf_nec_stableSNI2}
 \begin{aligned}
  p(s) = 
     \begin{cases}
        \frac{c\omega_0}{r(s^2+cs+\omega_0^2)} & \text{when }  \theta=\frac{\pi}{2}, \\
        \frac{(d^2-\omega_0^2)\operatorname{sec}\theta}{r\left(s^2+(d^2-\omega_0^2)\frac{\operatorname{tan}\theta}{\omega_0}s+d^2\right)} & \text{when }  \theta\in(0,\frac{\pi}{2}), \\
        \frac{(e^2-\omega_0^2)\operatorname{sec}\theta}{r\left(s^2+(e^2-\omega_0^2)\frac{\operatorname{tan}\theta}{\omega_0}s+e^2\right)} & \text{when }  \theta\in(\frac{\pi}{2},\pi).
     \end{cases} 
       \end{aligned}  
\end{equation}
   Evidently, $p(j\omega_0) x^* C(j\omega_0)x = 1$. Observe that $p(s)$ is SNI and has relative degree $2$. Let $P(s)=f(s) p(s) f(-s)^T$ and note that $P$ is SNI with $P(\infty) \geq 0$. Since, by construction, $[I-P(j\omega_0)C(j\omega_0)]x=0$, whereby $\det[I-P(j\omega_0)C(j\omega_0)] = 0$, it follows that $[P,C]$ is not stable. This shows by contraposition that $C$ must be NI.

   Second, we show that $C(0)\leq 0$. Suppose to the contrapositive that $C(0) \not\leq 0$. Since $C$ is NI, we know that $C(0) = C(0)^T \in \mathbb{R}^{n \times n}$ by Lemma~\ref{lem:propertyNI_zero_infty}(a) and at least one of its eigenvalues is greater than zero, whereby it admits a real Schur decomposition 
$C(0)=UDU^T$ with $U^T U = I$, $D$ being real diagonal and $[D]_{11} > 0$. Let $P=\frac{1}{s+1}M$ with
$$M = U\begin{bmatrix}
[D]_{11}^{-1} & \\
 & I_{n-1}
\end{bmatrix}U^T \geq 0.$$ Note that $P$ is SNI with $P(\infty) \geq 0$, and 
$$\det(I-P(0)C(0)) = 0,$$
whereby $[P,C]$ is unstable. By contraposition we know $C(0) \leq 0$. This completes the proof.
     $\hfill \qed$
\end{pf}

\black{We validate the correctness and demonstrate the usefulness of Theorem~\ref{thm: NI_nec} with the following example.  Suppose we wish to robustly stabilise \emph{all} SNI systems $P\in\mathcal{RH}_\infty$ satisfying $P(\infty)\geq 0$ with a constant feedback controller $C(s) = \alpha$, $\alpha\in\mathbb{R}$, then by Theorem~\ref{thm: NI_nec} we know that $\alpha \leq 0$ is necessary and sufficient. The sufficiency for robust stability can be easily verified using classical NI systems theory (e.g. Lemma~\ref{lem:stability_WithoutOriginPole2}). On the other hand, to more intuitively understand the necessity, instead of using the proof of Theorem~\ref{thm: NI_nec}, we note the following observations. Let a subset of the above SNI systems be characterised by
$$\mathcal{S}=\left\{P(s) = \frac{1}{s+\beta}:~\beta>0\right\}.$$
Then for any $P\in\mathcal{S}$, we have 
$$[1-P(s)C(s)]^{-1} = \frac{s+\beta}{s+\beta-\alpha},$$
which is stable if and only if $\beta-\alpha>0$. Since $\beta > 0$ is arbitrary, it follows that $[P,C]$ is robustly stable for all $P\in \mathcal{S}$ if and only if $\alpha \leq 0$. This thus demonstrates the necessity of the condition $\alpha \leq 0$ as well as the validity of Theorem~\ref{thm: NI_nec} for this specific example. 
}

Next, we also limit the static gain of the uncertain plant set and hence relax the static gain of the controller.

\begin{theorem} \label{thm: NI_nec_DC}
Let $C\in\mathcal{RH}_\infty^{n\times n}$ and $\gamma > 0$. Then $[P, C]$ is stable for all SNI $P \in \mathcal{RH}_\infty^{n \times n}$ satisfying $P(\infty) \ge 0$ and $P(0)  \leq \gamma I$ if and only if $C$ is NI and $C(0) < \frac{1}{\gamma} I$. 
\end{theorem}

\begin{pf}
(Sufficiency) The sufficiency proof follows by a similar routine to the proof for Theorem~\ref{thm: SNI_nec_DC_inst} with an application of Theorem~\ref{thm:stability_max_eigenvalue_sufficient_conditions}. 

 (Necessity)  First, we show that $C$ is NI via a contrapositive argument. Suppose $C$ is not NI. Then, $\exists\,\omega_0\in(0,\infty), x\in\mathbb{C}^{n}\backslash\{0\}$ such that $x^* j[C(j\omega_0)-C(j\omega_0)^*]x< 0$. This equivalent to $\Im(x^* C(j\omega_0) x) > 0$. Choose $\alpha,\beta\in\mathbb{R}^n$ such that $\alpha j\omega_0 +\beta = x$, and define $f(s)=(\alpha s + \beta) \in\mathbb{C}^n$. 
Choose $r>0$ and $\theta\in(0,\pi)$ such that $r e^{j\theta}=x^* C(j\omega_0) x$. Let $c,d,e,p(s)$ be defined in \eqref{eq:pf_nec_stableSNI1} and \eqref{eq:pf_nec_stableSNI2}. 
Clearly, $p(s)$ is SNI and has relative degree $2$. Let $P(s)=f(s) p(s) f(-s)^T$ and it holds that $P$ is SNI with $P(\infty) \geq 0$. Additionally, one may verify that $P(0) \leq \gamma I$. Since, by construction, $[I-P(j\omega_0)C(j\omega_0)]x=0$, whereby $\det[I-P(j\omega_0)C(j\omega_0)] = 0$, it follows that $[P,C]$ is not stable. This shows by contraposition that $C$ must be NI. 

 Second, similarly to the proof of Theorem~\ref{thm: SNI_nec_DC_inst}, we show that $C(0) < \frac{1}{\gamma} I$ via a contrapositive argument. Suppose that  $C(0) \not < \frac{1}{\gamma} I$. Then, $\overline{\lambda}[C(0)]\geq \frac{1}{\gamma}$. Choose $P(s)=\frac{1}{\overline{\lambda}[C(0)] (s+1)} I$ and note that $P$ is SNI, which satisfies $P(0) \leq \gamma I$ and $P(\infty) \geq 0$. Since, by construction, $\det[I-P(0)C(0)] = 0$, it follows that $[P,C]$ is not stable. This shows by contraposition that $C(0) < \frac{1}{\gamma} I$.
    $\hfill \qed$
\end{pf}

Finally, we remove the restriction on the instantaneous gains of the uncertain plant set and instead impose the corresponding restriction on the instantaneous gain of the controller.

\begin{theorem} \label{thm: NI_nec_DC_inst}
Let $C\in\mathcal{RH}_\infty^{n\times n}$ and $\gamma > 0$.
%Then $[P, C]$ is stable for all SNI $P \in \mathcal{RH}_\infty^{n \times n}$ satisfying $P(0) \leq \gamma I$ if and only if $C$ is NI, $C(\infty) \ge 0$, and $C(0) < \frac{1}{\gamma} I$. 
\black{Then the following three statements are equivalent:
\begin{enumerate} \renewcommand{\theenumi}{\textup{(\roman{enumi})}}\renewcommand{\labelenumi}{\theenumi}
    \item $[P, C]$ is stable for all SNI $P \in \mathcal{RH}_\infty^{n \times n}$ satisfying $P(0) \leq \gamma I$; \label{item: NI_nec_DC_inst_1}
    \item $[P, C]$ is stable for all SNI $P \in \mathcal{RH}_\infty^{n \times n}$ satisfying $0 \leq P(0) \leq \gamma I$; \label{item: NI_nec_DC_inst_2}   %$P\in\mathcal{N}_1$; 
    \item $C$ is NI, $C(\infty) \ge 0$, and $C(0) < \frac{1}{\gamma} I$. \label{item: NI_nec_DC_inst_C}
\end{enumerate}
}
\end{theorem}

\begin{pf}
\black{That \ref{item: NI_nec_DC_inst_1} implies \ref{item: NI_nec_DC_inst_2} is trivial. That \ref{item: NI_nec_DC_inst_C} implies  \ref{item: NI_nec_DC_inst_1}}  can be shown by a similar routine to the proof for Theorem~\ref{thm: SNI_nec_DC_inst} with an application of Theorem~\ref{thm:stability_max_eigenvalue_sufficient_conditions}. 

\black{We next show that \ref{item: NI_nec_DC_inst_2} implies \ref{item: NI_nec_DC_inst_C} in what follows.} By Theorem~\ref{thm: NI_nec_DC}, $C$ is NI and $C(0) < \frac{1}{\gamma} I$. It thus suffices to show that $C(\infty)\geq 0$. Suppose to the contrapositive that $C(\infty)\not\geq 0$. Then, $\underline{\lambda}[C(\infty)]<0$. Choose
  $P(s) = \frac{s}{\underline{\lambda}[C(\infty)](s+1)}I_n$. Note that $P$ is SNI and satisfies \black{$0\leq P(0) \leq\gamma I$}. Since, by construction, $\det[I-P(\infty)C(\infty)] = 0$, it follows that $[P,C]$ is not stable. This shows by contraposition that $C(\infty) \geq 0$, which completes the proof. 
    $\hfill \qed$
\end{pf}

%\subsection{Conjecture}

% \begin{conjecture} \label{thm: OSNI_nec_DC}
% Let $C\in\mathcal{RH}_\infty^{n\times n}$ and $\gamma > 0$. Then $[P, C]$ is stable for all NI $P \in \mathcal{RH}_\infty^{n \times n}$ satisfying $P(\infty) = 0$ and $P(0) \leq \gamma I$ if and only if $C$ is OSNI and $C(0) < \frac{1}{\gamma}{I}$. 
% \end{conjecture}

% \begin{pf}
% (Sufficiency) 
% % The sufficiency follows by a similar routine as that in the proof for Theorem~\ref{thm: OSNI_nec}. 
% % Let $\Pi_0 = \begin{bmatrix}
% %      -\delta I & -I\\
% %      -I & 2\epsilon I
% %  \end{bmatrix}$

% (Necessity) 

% We first show $C$ is OSNI. Suppose to the contrapositive that for some $\omega_0\in(0,\infty)$, there exists no $\epsilon>0$ for which \eqref{eq:pf_step1} holds. Use the same notation and construction in the proof for Theorem~\ref{thm: OSNI_nec}. One can verify that 
% \begin{align*}
% P(0) &= \lim_{s\to 0} \frac{\omega_0}{2s}(I+G(s)) \\
% &=\lim_{s\to 0} \frac{\omega_0}{2s}(g_1(s)Q+(1-g_2(s))I)\\
% &=\lim_{s\to 0} \frac{\omega_0}{2s}\left(\frac{\delta s}{s^2+\delta s+\omega_0^2}Q + \frac{\omega_0^2(s+\delta)^2-\delta^2(s^2+\omega_0^2)}{\omega_0^2(s+\delta)^2}I\right)\\
% &= \frac{\omega_0}{\delta}I+\frac{\delta}{2\omega_0}Q
% \end{align*}

%     $\hfill \qed$
% \end{pf}

\section{Conclusions} \label{sec: con}

Robust feedback stability against various uncertainty classes of NI systems was studied. It was shown that there exists no stable controller that may stabilise all NI plants possibly with marginally stable poles on the imaginary axis. In other words, the uncertainty set of all such NI plants is overly large for any strongly stabilising controller to exist. By constraining the static (i.e. $\omega = 0$) and instantaneous (i.e. $\omega = \infty$) gains of the uncertain NI \black{or SNI} plants, it was established that in order for a controller to be robustly stabilising, it must exhibit a certain type of NI property. 

The converse results obtained in this paper signify that the NI property is nonconservative within the context of robustly stabilising NI uncertain systems. Since NI uncertainty naturally arises in various physical systems with colocated force actuators and displacement sensors, these results motivate and justify NI controller synthesis, a future research direction of great interest and practical values.

\bibliographystyle{apacite}  
\bibliography{Neg_Img}

\end{document}